\let\latexdocument\document
\let\latexarabic\arabic

\documentclass[lineno]{biometrika}

\usepackage{amsmath,amsfonts,amssymb}
\usepackage{tikz}

\usepackage{times}
\usepackage{bm}
\usepackage{natbib}
\usepackage{enumerate}
\usepackage{color}

\newcommand{\E}{\mathbb{E}}
\newcommand{\Prob}{\mathbb{P}}
\newcommand{\V}{\mathbb{V}}
\newcommand{\Cov}{\text{Cov}}

\newcommand\independent{\protect\mathpalette{\protect\independenT}{\perp}}
\def\independenT#1#2{\mathrel{\rlap{$#1#2$}\mkern2mu{#1#2}}}

\let\document\latexdocument
\let\arabic\latexarabic

\def\rm{}

\begin{document}

\jname{}
%% The year, volume, and number are determined on publication
\jyear{Research Note}
\jvol{}
\jnum{}

%% Here are the title, author names and addresses
\title{On true versus estimated propensity scores for treatment effect estimation with discrete controls}

\author{A. Herren and P.R. Hahn}
\affil{School of Mathematical and Statistical Sciences, Arizona State University,\\ Tempe, AZ 85281, U.S.
\email{asherren@asu.edu} \email{prhahn@asu.edu}}

\maketitle

\begin{abstract}
The finite sample variance of an inverse propensity weighted estimator is derived in the case of discrete control variables with finite support. The obtained expressions generally corroborate widely-cited asymptotic theory showing that estimated propensity scores are superior to true propensity scores in the context of inverse propensity weighting. However, similar analysis of a modified estimator demonstrates that foreknowledge of the true propensity function can confer a statistical advantage when estimating average treatment effects. 
\end{abstract}

\section{Introduction}
\subsection{Background}
\cite{rosenbaum1983central} showed that if a set of covariates satisfy conditional unconfoundedness, the univariate propensity score based on these controls (the conditional expectation of treatment assignment), also satisfies conditional unconfoundedness.  Several classic papers have since shown that \textit{even when the true propensity function is known}, the treatment effect can be estimated with greater efficiency using nonparametric estimates of the propensity function (\cite{robins1992estimating}; \cite{hahn1998role}; \cite{hirano2003efficient}). 
This paper derives a finite-sample analogue of these results in the case of discrete control variables with finite support. On its face, these results would seem to imply that, with a suitably flexible model of treatment given controls, knowledge of the true propensity function is useless. While this is narrowly true for inverse-propensity weighting (IPW), variance analysis of a modified IPW shows that incorporating knowledge of the true propensity score can improve statistical precision in some cases.

\subsection{Notation}

Let $Z$ denote a binary treatment, $X$ denote a discrete covariate taking $k$ unique values, $\mathcal{X} = \lbrace 1, \dots, k \rbrace$, and $Y$ denote the outcome of interest. 
We are interested in the average causal effect of $Z$ on $Y$, which can be expressed in terms of two counterfactual random variables, 
$Y^1$ and $Y^0$, the outcomes in which the treatment $Z$ is set to 1 and 0, respectively. 
We make several standard identifying assumptions:
\begin{itemize}
\item Stable unit treatment value assumption (SUTVA)
\begin{itemize}
\item \textit{Consistency}: $Y = Y^1Z + Y^0(1-Z)$
\item \textit{No Interference}: For any two indices $1 \leq i, j \leq n$ in a dataset of size $n$, $(Y^1_i, Y^0_i) \independent Z_j$
\end{itemize}
\item Positivity: $0 < \Prob\left(Z=1 \mid X = x\right) < 1$ for all $x \in \mathcal{X}$
\item Conditional unconfoundedness: $(Y^1, Y^0) \independent Z \mid X$
\end{itemize}
These assumptions allow us to identify the average treatment effect (ATE) as an estimable contrast:
\begin{equation*}
\E[Y^1 - Y^0] = \E_{X}[\E[Y \mid X, Z = 1] - \E[Y \mid X, Z = 0]].
\end{equation*}
The propensity score, $p(X) = \mathbb{P}(Z = 1 \mid X)$, likewise satisfies conditional unconfoundedness:
\begin{equation*}
\E[Y^1 - Y^0] = \E_{p(X)}[\E[Y \mid p(X), Z = 1] - \E[Y \mid p(X), Z = 0]].
\end{equation*}

Many estimators target this estimand; in this paper we focus on inverse-propensity weight estimators for arbitrary weighting function $w$:
\begin{equation} \label{ipw}
\bar{\tau}_{\textrm{ipw}}^w = \frac{1}{N} \sum_{i=1}^{N} \left( \frac{Y_i Z_i}{w(X_i)} - \frac{Y_i (1-Z_i)}{1-w(X_i)} \right).
\end{equation}
Note that the weight function itself may be estimated from the sample data.

\section{Variance comparison of true versus estimated IPW}\label{trueipw}
We assume throughout that every strata of $\mathcal{X}$ has at least one treated-control pair: $N_{x,1} > 0$ and $N_{x,0} > 0$ for all $x$. Direct calculation reveals that for a putative propensity function $w: \mathcal{X} \mapsto (0, 1)$, the corresponding IPW estimator may be written as
\begin{equation}\label{strat_sum}
\bar{\tau}_{\textrm{ipw}}^w = \sum_x \left( \frac{N_x}{N} \right) \bar{\tau}_{\textrm{ipw}}^{w,x}
\end{equation}
where
\begin{equation}\label{strat_contrast}
\bar{\tau}_{\textrm{ipw}}^{w,x} = \left( \frac{\hat{p}(x)}{w(x)} \bar{Y}_{x, Z=1} - \frac{1-\hat{p}(x)}{1-w(x)} \bar{Y}_{x, Z=0} \right)
\end{equation}
and  $\hat{p}(x) = \left( N_{x, Z=1} / N_{x} \right)$ is the proportion of treated units in each stratum.

Taking $w(x) = p(x)$ is the ``true propensity score'' case, while letting $w(x) = \hat{p}(x)$ is the ``estimated propensity score'' case. In the former case, the treated and untreated stratum averages are weighted by $\hat{p}(x) / p(x)$ and $\hat{q}(x) / q(x) = \left( 1-\hat{p}(x) \right) / \left(1-p(x) \right)$, respectively; in the latter case the weights are identically one. This difference in weights leads to the following analogue of the results of \cite{hahn1998role}, \cite{hirano2003efficient}, and \cite{robins1992estimating}:

\begin{proposition} \label{theorem1}
The variance difference between the true propensity score IPW and the estimated propensity score IPW may be expressed as
\small
\begin{equation}
\begin{split}
\sum_{x \in \mathcal{X}} \V(\bar{\tau}_{\textrm{ipw}}^{p, x}) - \V(\bar{\tau}_{\textrm{ipw}}^{\hat{p},x})  &= \sum_{x \in \mathcal{X}}  \left ( \frac{\mu_1(x)}{p(x)} + \frac{\mu_0(x)}{1-p(x)} \right )^2 \frac{n_xp(x)(1 - p(x))}{ (n_x + 2)^2}\\
 & \;\;\;\;\; + \sum_{x \in \mathcal{X}} \sigma^2_{x,1}\left( \frac{p(x)n_x + 1}{(n_x+2)^2p(x)^2} -  \frac{1- (1 - p(x))^{(n_x+1)}}{(n_x+1)p(x)} \right)\\
& \;\;\;\;\; + \sum_{x \in \mathcal{X}}   \sigma^2_{x,0} \left (\frac{(1- p(x))n_x + 1}{(n_x+2)^2(1 - p(x))^2}  - \frac{1- p(x)^{(n_x+1)}}{(n_x+1)(1 - p(x))} \right)
\end{split}
\end{equation}
\normalsize
where 
$n_x = N_x - 2$.
\end{proposition}

We arrive at this expression by deriving it for an individual stratum, suppressing the dependence on $x$ for notational convenience. By the Law of Total Variance
\small
\begin{equation}\label{LTV}
\begin{split}
\V  \left( \frac{\hat{p}}{w} \bar{Y}_{1} - \frac{1-\hat{p}}{1-w} \bar{Y}_{0} \right) &= \V  \left(  \E \left (\frac{\hat{p}}{w} \bar{Y}_{1} - \frac{1-\hat{p}}{1-w} \bar{Y}_{0} \mid Z \right) \right) + \E \left( \V \left (\frac{\hat{p}}{w} \bar{Y}_{1} - \frac{1-\hat{p}}{1-w} \bar{Y}_{0} \mid Z  \right) \right)\\
&=  \V  \left(  \frac{\hat{p}}{w} \mu_1 - \frac{1-\hat{p}}{1-w} \mu_0 \right) + \E \left(\left(\frac{\hat{p}}{w}\right)^2 \frac{\sigma^2_1}{N_1} + \left(\frac{1-\hat{p}}{1-w}\right)^2 \frac{\sigma^2_0}{N_0}  \right).
\end{split}
\end{equation}
\normalsize
The random variable in these expressions is $\hat{p} =  N_1 / N$ and $\hat{q} = 1-\hat{p} =  N_0 / N$, respectively. We handle each term separately.

When $w = N_1/N$ (itself a random variable), as in the estimated IPW, then the first term becomes 0 (the variance of a constant). Meanwhile, when the weighting function is fixed at $w = p$ (and $1-w = 1 - p = q$) we find that
\small
\begin{equation}\label{quadform}
\begin{split}
 \V  \left(  \frac{\hat{p}}{p} \mu_1 - \frac{1-\hat{p}}{1-p} \mu_0 \right) &= \frac{\mu_1^2}{p^2 N^2}\V (N_1) + \frac{\mu_0^2}{q^2 N^2} \V(N_0) - 2 \frac{\mu_1 \mu_0}{pqN^2} \Cov\left(N_1, N_0\right)\\
 &= \left ( \frac{\mu_1^2}{p^2 N^2}+ \frac{\mu_0^2}{q^2 N^2}  + 2 \frac{\mu_1 \mu_0}{pqN^2} \right ) n p q,\\
 &= \left ( \frac{\mu_1}{p} + \frac{\mu_0}{q }  \right )^2 \frac{n p q}{(n+2)^2},
 \end{split}
\end{equation}
\normalsize
for $n = N - 2$ (because our observations are constrained to have at least one treated and untreated unit) and using the fact that $N_1 = N - N_0$ and therefore $\Cov\left(N_1, N_0\right) = -\V(N_1) = -\V(N_0)$.

On to the second term, we consider 
\small
\begin{equation*}
 \E \left( \frac{\sigma^2_1}{N_1} +  \frac{\sigma^2_0}{N_0}  \right)
 \end{equation*}\normalsize
and
\small
\begin{equation*}
 \E \left(\left(\frac{\hat{p}}{p}\right)^2 \frac{\sigma^2_1}{N_1} + \left(\frac{1-\hat{p}}{1-p}\right)^2 \frac{\sigma^2_0}{N_0}  \right)
 \end{equation*}
\normalsize
in turn. We apply, in each case, a result of \cite{chao1972negative} about negative moments of a positive random variable $A$; that is, expectations of the form $\E \left (1/(c + A) \right)$ for $c$ a known constant. Specifically, for $A \sim \mbox{Bin}(n, p)$ and $c = 1$, they show that  
$$\E \left ( \frac{1}{(1 + A)} \right ) = \frac{1 - q^{(n+1)}}{(n + 1) p}. $$ We apply this result to the estimated propensity IPW by writing $N_1 = 1 + A$ where $A$ has parameters $n = N - 2$ and $p$. The number of observed controls, $N_0$ has a similar representation, but with parameter $q = 1 - p$. Therefore
\begin{equation}
 \E \left( \frac{\sigma^2_1}{N_1} +  \frac{\sigma^2_0}{N_0}  \right) =  \sigma_1^2 \frac{1- q^{(n+1)}}{(n+1)p} +   \sigma_0^2 \frac{1- p^{(n+1)}}{(n+1)q}.
\end{equation}
The true propensity IPW is more straightforward, but uses the same representation of the treated (respectively, control) counts as $1 + A$ for a binomial random variable with parameters $n = N - 2$ and $p$ (respectively, $q = 1 - p$):
\begin{equation}
\begin{split}
 \E \left(\left(\frac{\hat{p}}{p}\right)^2 \frac{\sigma^2_1}{N_1} + \left(\frac{1-\hat{p}}{1-p}\right)^2 \frac{\sigma^2_0}{N_0}  \right) &= \frac{\sigma_1^2}{N^2 p^2}\E(N_1) + \frac{\sigma_1^2}{N^2 (1 - p)^2}\E(N_0)\\
 &= \sigma_1^2 \frac{ (p n + 1 )}{(n+2)^2 p^2} + \sigma_0^2 \frac{ (q n + 1 )}{(n+2)^2 q^2}.
\end{split}
 \end{equation}
Gathering like terms and summing over strata, indexed by $x$, demonstrates the proposition.

\begin{figure}\caption{A comparison of the finite sample variances when $N_x = 17$, with $\sigma_1^2 = 4$ and $\sigma^2_0 = 16$. Left panel: $\mu_1 = \mu_0 = 0$. Right panel:  $\mu_1 = 1$ and $\mu_0 = 3$.}\label{comparison}
\centering

\begin{tikzpicture}[x=1pt,y=1pt]
\definecolor{fillColor}{RGB}{255,255,255}
\path[use as bounding box,fill opacity=0.00] (0,0) rectangle (397.48,180.67);
\begin{scope}
\path[clip] ( 49.20, 54.00) rectangle (192.74,167.47);
\definecolor{drawColor}{RGB}{0,0,0}

\path[draw=drawColor,line width= 0.4pt,line join=round,line cap=round] ( 55.18,123.01) --
	( 55.85,121.35) --
	( 56.52,119.76) --
	( 57.19,118.25) --
	( 57.86,116.80) --
	( 58.52,115.42) --
	( 59.19,114.10) --
	( 59.86,112.84) --
	( 60.53,111.64) --
	( 61.20,110.49) --
	( 61.86,109.39) --
	( 62.53,108.34) --
	( 63.20,107.34) --
	( 63.87,106.39) --
	( 64.53,105.48) --
	( 65.20,104.61) --
	( 65.87,103.78) --
	( 66.54,102.99) --
	( 67.21,102.23) --
	( 67.87,101.51) --
	( 68.54,100.82) --
	( 69.21,100.17) --
	( 69.88, 99.54) --
	( 70.55, 98.95) --
	( 71.21, 98.38) --
	( 71.88, 97.84) --
	( 72.55, 97.33) --
	( 73.22, 96.84) --
	( 73.89, 96.37) --
	( 74.55, 95.93) --
	( 75.22, 95.51) --
	( 75.89, 95.11) --
	( 76.56, 94.73) --
	( 77.22, 94.36) --
	( 77.89, 94.02) --
	( 78.56, 93.70) --
	( 79.23, 93.39) --
	( 79.90, 93.10) --
	( 80.56, 92.82) --
	( 81.23, 92.56) --
	( 81.90, 92.31) --
	( 82.57, 92.08) --
	( 83.24, 91.86) --
	( 83.90, 91.66) --
	( 84.57, 91.46) --
	( 85.24, 91.28) --
	( 85.91, 91.11) --
	( 86.58, 90.95) --
	( 87.24, 90.81) --
	( 87.91, 90.67) --
	( 88.58, 90.54) --
	( 89.25, 90.43) --
	( 89.91, 90.32) --
	( 90.58, 90.22) --
	( 91.25, 90.14) --
	( 91.92, 90.06) --
	( 92.59, 89.99) --
	( 93.25, 89.92) --
	( 93.92, 89.87) --
	( 94.59, 89.82) --
	( 95.26, 89.78) --
	( 95.93, 89.75) --
	( 96.59, 89.73) --
	( 97.26, 89.71) --
	( 97.93, 89.70) --
	( 98.60, 89.70) --
	( 99.26, 89.70) --
	( 99.93, 89.71) --
	(100.60, 89.73) --
	(101.27, 89.76) --
	(101.94, 89.79) --
	(102.60, 89.82) --
	(103.27, 89.87) --
	(103.94, 89.92) --
	(104.61, 89.97) --
	(105.28, 90.03) --
	(105.94, 90.10) --
	(106.61, 90.17) --
	(107.28, 90.25) --
	(107.95, 90.34) --
	(108.62, 90.43) --
	(109.28, 90.53) --
	(109.95, 90.63) --
	(110.62, 90.74) --
	(111.29, 90.86) --
	(111.95, 90.98) --
	(112.62, 91.10) --
	(113.29, 91.24) --
	(113.96, 91.38) --
	(114.63, 91.52) --
	(115.29, 91.67) --
	(115.96, 91.83) --
	(116.63, 92.00) --
	(117.30, 92.17) --
	(117.97, 92.34) --
	(118.63, 92.53) --
	(119.30, 92.72) --
	(119.97, 92.92) --
	(120.64, 93.12) --
	(121.31, 93.33) --
	(121.97, 93.55) --
	(122.64, 93.77) --
	(123.31, 94.01) --
	(123.98, 94.25) --
	(124.64, 94.50) --
	(125.31, 94.75) --
	(125.98, 95.02) --
	(126.65, 95.29) --
	(127.32, 95.57) --
	(127.98, 95.86) --
	(128.65, 96.16) --
	(129.32, 96.46) --
	(129.99, 96.78) --
	(130.66, 97.11) --
	(131.32, 97.44) --
	(131.99, 97.79) --
	(132.66, 98.15) --
	(133.33, 98.51) --
	(134.00, 98.89) --
	(134.66, 99.28) --
	(135.33, 99.68) --
	(136.00,100.10) --
	(136.67,100.52) --
	(137.33,100.96) --
	(138.00,101.42) --
	(138.67,101.88) --
	(139.34,102.36) --
	(140.01,102.86) --
	(140.67,103.37) --
	(141.34,103.90) --
	(142.01,104.44) --
	(142.68,105.00) --
	(143.35,105.58) --
	(144.01,106.18) --
	(144.68,106.80) --
	(145.35,107.43) --
	(146.02,108.09) --
	(146.68,108.77) --
	(147.35,109.47) --
	(148.02,110.19) --
	(148.69,110.94) --
	(149.36,111.71) --
	(150.02,112.51) --
	(150.69,113.33) --
	(151.36,114.19) --
	(152.03,115.07) --
	(152.70,115.99) --
	(153.36,116.93) --
	(154.03,117.91) --
	(154.70,118.92) --
	(155.37,119.97) --
	(156.04,121.06) --
	(156.70,122.19) --
	(157.37,123.36) --
	(158.04,124.57) --
	(158.71,125.82) --
	(159.37,127.12) --
	(160.04,128.47) --
	(160.71,129.88) --
	(161.38,131.33) --
	(162.05,132.84) --
	(162.71,134.41) --
	(163.38,136.04) --
	(164.05,137.73) --
	(164.72,139.49) --
	(165.39,141.31) --
	(166.05,143.21) --
	(166.72,145.19) --
	(167.39,147.24) --
	(168.06,149.38) --
	(168.73,151.61) --
	(169.39,153.92) --
	(170.06,156.33) --
	(170.73,158.84) --
	(171.40,161.45) --
	(172.06,164.17) --
	(172.73,167.00) --
	(173.40,169.95) --
	(174.07,173.03) --
	(174.74,176.23) --
	(175.40,179.58) --
	(176.07,183.06) --
	(176.74,186.69) --
	(177.41,190.48) --
	(178.08,194.43) --
	(178.74,198.56) --
	(179.41,202.86) --
	(180.08,207.34) --
	(180.75,212.03) --
	(181.42,216.92) --
	(182.08,222.02) --
	(182.75,227.34) --
	(183.42,232.91) --
	(184.09,238.71) --
	(184.75,244.78) --
	(185.42,251.11) --
	(186.09,257.72) --
	(186.76,264.63);
\end{scope}
\begin{scope}
\path[clip] (  0.00,  0.00) rectangle (397.48,180.67);
\definecolor{drawColor}{RGB}{0,0,0}

\path[draw=drawColor,line width= 0.4pt,line join=round,line cap=round] ( 54.52, 54.00) -- (187.43, 54.00);

\path[draw=drawColor,line width= 0.4pt,line join=round,line cap=round] ( 54.52, 54.00) -- ( 54.52, 48.00);

\path[draw=drawColor,line width= 0.4pt,line join=round,line cap=round] ( 81.10, 54.00) -- ( 81.10, 48.00);

\path[draw=drawColor,line width= 0.4pt,line join=round,line cap=round] (107.68, 54.00) -- (107.68, 48.00);

\path[draw=drawColor,line width= 0.4pt,line join=round,line cap=round] (134.26, 54.00) -- (134.26, 48.00);

\path[draw=drawColor,line width= 0.4pt,line join=round,line cap=round] (160.84, 54.00) -- (160.84, 48.00);

\path[draw=drawColor,line width= 0.4pt,line join=round,line cap=round] (187.43, 54.00) -- (187.43, 48.00);

\node[text=drawColor,anchor=base,inner sep=0pt, outer sep=0pt, scale=  1.00] at ( 54.52, 32.40) {0.0};

\node[text=drawColor,anchor=base,inner sep=0pt, outer sep=0pt, scale=  1.00] at ( 81.10, 32.40) {0.2};

\node[text=drawColor,anchor=base,inner sep=0pt, outer sep=0pt, scale=  1.00] at (107.68, 32.40) {0.4};

\node[text=drawColor,anchor=base,inner sep=0pt, outer sep=0pt, scale=  1.00] at (134.26, 32.40) {0.6};

\node[text=drawColor,anchor=base,inner sep=0pt, outer sep=0pt, scale=  1.00] at (160.84, 32.40) {0.8};

\node[text=drawColor,anchor=base,inner sep=0pt, outer sep=0pt, scale=  1.00] at (187.43, 32.40) {1.0};

\path[draw=drawColor,line width= 0.4pt,line join=round,line cap=round] ( 49.20, 58.20) -- ( 49.20,163.27);

\path[draw=drawColor,line width= 0.4pt,line join=round,line cap=round] ( 49.20, 58.20) -- ( 43.20, 58.20);

\path[draw=drawColor,line width= 0.4pt,line join=round,line cap=round] ( 49.20, 84.47) -- ( 43.20, 84.47);

\path[draw=drawColor,line width= 0.4pt,line join=round,line cap=round] ( 49.20,110.74) -- ( 43.20,110.74);

\path[draw=drawColor,line width= 0.4pt,line join=round,line cap=round] ( 49.20,137.00) -- ( 43.20,137.00);

\path[draw=drawColor,line width= 0.4pt,line join=round,line cap=round] ( 49.20,163.27) -- ( 43.20,163.27);

\node[text=drawColor,rotate= 90.00,anchor=base,inner sep=0pt, outer sep=0pt, scale=  1.00] at ( 34.80, 58.20) {0};

\node[text=drawColor,rotate= 90.00,anchor=base,inner sep=0pt, outer sep=0pt, scale=  1.00] at ( 34.80, 84.47) {2};

\node[text=drawColor,rotate= 90.00,anchor=base,inner sep=0pt, outer sep=0pt, scale=  1.00] at ( 34.80,110.74) {4};

\node[text=drawColor,rotate= 90.00,anchor=base,inner sep=0pt, outer sep=0pt, scale=  1.00] at ( 34.80,137.00) {6};

\node[text=drawColor,rotate= 90.00,anchor=base,inner sep=0pt, outer sep=0pt, scale=  1.00] at ( 34.80,163.27) {8};
\end{scope}
\begin{scope}
\path[clip] (  0.00,  0.00) rectangle (198.74,180.67);
\definecolor{drawColor}{RGB}{0,0,0}

\node[text=drawColor,anchor=base,inner sep=0pt, outer sep=0pt, scale=  1.00] at (120.97,  8.40) {Propensity};

\node[text=drawColor,rotate= 90.00,anchor=base,inner sep=0pt, outer sep=0pt, scale=  1.00] at ( 10.80,110.74) {Variance};
\end{scope}
\begin{scope}
\path[clip] ( 49.20, 54.00) rectangle (192.74,167.47);
\definecolor{drawColor}{RGB}{0,0,0}

\path[draw=drawColor,line width= 0.4pt,dash pattern=on 4pt off 4pt ,line join=round,line cap=round] ( 56.19,1770.61) --
	( 56.52,1164.26) --
	( 57.19,721.63) --
	( 57.86,510.27) --
	( 58.52,391.95) --
	( 59.19,318.52) --
	( 59.86,269.51) --
	( 60.53,234.99) --
	( 61.20,209.65) --
	( 61.86,190.42) --
	( 62.53,175.44) --
	( 63.20,163.50) --
	( 63.87,153.81) --
	( 64.53,145.83) --
	( 65.20,139.15) --
	( 65.87,133.50) --
	( 66.54,128.67) --
	( 67.21,124.51) --
	( 67.87,120.90) --
	( 68.54,117.73) --
	( 69.21,114.94) --
	( 69.88,112.47) --
	( 70.55,110.26) --
	( 71.21,108.29) --
	( 71.88,106.52) --
	( 72.55,104.92) --
	( 73.22,103.47) --
	( 73.89,102.16) --
	( 74.55,100.96) --
	( 75.22, 99.87) --
	( 75.89, 98.87) --
	( 76.56, 97.95) --
	( 77.22, 97.10) --
	( 77.89, 96.32) --
	( 78.56, 95.61) --
	( 79.23, 94.95) --
	( 79.90, 94.33) --
	( 80.56, 93.77) --
	( 81.23, 93.24) --
	( 81.90, 92.76) --
	( 82.57, 92.31) --
	( 83.24, 91.89) --
	( 83.90, 91.50) --
	( 84.57, 91.14) --
	( 85.24, 90.81) --
	( 85.91, 90.50) --
	( 86.58, 90.21) --
	( 87.24, 89.95) --
	( 87.91, 89.70) --
	( 88.58, 89.47) --
	( 89.25, 89.26) --
	( 89.91, 89.07) --
	( 90.58, 88.90) --
	( 91.25, 88.73) --
	( 91.92, 88.59) --
	( 92.59, 88.45) --
	( 93.25, 88.33) --
	( 93.92, 88.22) --
	( 94.59, 88.12) --
	( 95.26, 88.04) --
	( 95.93, 87.96) --
	( 96.59, 87.90) --
	( 97.26, 87.84) --
	( 97.93, 87.80) --
	( 98.60, 87.76) --
	( 99.26, 87.74) --
	( 99.93, 87.72) --
	(100.60, 87.71) --
	(101.27, 87.71) --
	(101.94, 87.72) --
	(102.60, 87.73) --
	(103.27, 87.76) --
	(103.94, 87.79) --
	(104.61, 87.83) --
	(105.28, 87.87) --
	(105.94, 87.93) --
	(106.61, 87.99) --
	(107.28, 88.06) --
	(107.95, 88.13) --
	(108.62, 88.21) --
	(109.28, 88.30) --
	(109.95, 88.40) --
	(110.62, 88.50) --
	(111.29, 88.62) --
	(111.95, 88.73) --
	(112.62, 88.86) --
	(113.29, 88.99) --
	(113.96, 89.13) --
	(114.63, 89.28) --
	(115.29, 89.43) --
	(115.96, 89.59) --
	(116.63, 89.76) --
	(117.30, 89.93) --
	(117.97, 90.12) --
	(118.63, 90.31) --
	(119.30, 90.51) --
	(119.97, 90.71) --
	(120.64, 90.93) --
	(121.31, 91.15) --
	(121.97, 91.38) --
	(122.64, 91.62) --
	(123.31, 91.87) --
	(123.98, 92.13) --
	(124.64, 92.39) --
	(125.31, 92.67) --
	(125.98, 92.95) --
	(126.65, 93.25) --
	(127.32, 93.56) --
	(127.98, 93.87) --
	(128.65, 94.20) --
	(129.32, 94.54) --
	(129.99, 94.89) --
	(130.66, 95.25) --
	(131.32, 95.62) --
	(131.99, 96.01) --
	(132.66, 96.41) --
	(133.33, 96.82) --
	(134.00, 97.25) --
	(134.66, 97.69) --
	(135.33, 98.15) --
	(136.00, 98.62) --
	(136.67, 99.11) --
	(137.33, 99.61) --
	(138.00,100.14) --
	(138.67,100.68) --
	(139.34,101.24) --
	(140.01,101.82) --
	(140.67,102.43) --
	(141.34,103.05) --
	(142.01,103.70) --
	(142.68,104.37) --
	(143.35,105.07) --
	(144.01,105.80) --
	(144.68,106.55) --
	(145.35,107.34) --
	(146.02,108.15) --
	(146.68,109.00) --
	(147.35,109.88) --
	(148.02,110.80) --
	(148.69,111.76) --
	(149.36,112.76) --
	(150.02,113.80) --
	(150.69,114.89) --
	(151.36,116.03) --
	(152.03,117.22) --
	(152.70,118.47) --
	(153.36,119.78) --
	(154.03,121.15) --
	(154.70,122.59) --
	(155.37,124.10) --
	(156.04,125.69) --
	(156.70,127.36) --
	(157.37,129.13) --
	(158.04,131.00) --
	(158.71,132.97) --
	(159.37,135.05) --
	(160.04,137.26) --
	(160.71,139.61) --
	(161.38,142.10) --
	(162.05,144.76) --
	(162.71,147.59) --
	(163.38,150.62) --
	(164.05,153.87) --
	(164.72,157.35) --
	(165.39,161.09) --
	(166.05,165.13) --
	(166.72,169.49) --
	(167.39,174.22) --
	(168.06,179.36) --
	(168.73,184.96) --
	(169.39,191.09) --
	(170.06,197.82) --
	(170.73,205.24) --
	(171.40,213.46) --
	(172.06,222.59) --
	(172.73,232.80) --
	(173.40,244.28) --
	(174.07,257.26) --
	(174.74,272.04) --
	(175.40,288.99) --
	(176.07,308.59) --
	(176.74,331.48) --
	(177.41,358.48) --
	(178.08,390.73) --
	(178.74,429.77) --
	(179.41,477.81) --
	(180.08,538.02) --
	(180.75,615.21) --
	(181.42,716.85) --
	(182.08,855.20) --
	(182.75,1051.50) --
	(183.42,1345.50) --
	(184.02,1770.61);

\path[draw=drawColor,line width= 0.4pt,line join=round,line cap=round] (107.68, 81.19) rectangle (168.17, 59.59);

\path[draw=drawColor,line width= 0.4pt,line join=round,line cap=round] (113.08, 73.99) -- (123.88, 73.99);

\path[draw=drawColor,line width= 0.4pt,dash pattern=on 4pt off 4pt ,line join=round,line cap=round] (113.08, 66.79) -- (123.88, 66.79);

\node[text=drawColor,anchor=base west,inner sep=0pt, outer sep=0pt, scale=  0.60] at (129.28, 71.92) {Estimated-PS};

\node[text=drawColor,anchor=base west,inner sep=0pt, outer sep=0pt, scale=  0.60] at (129.28, 64.72) {True-PS};
\end{scope}
\begin{scope}
\path[clip] (247.94, 54.00) rectangle (391.48,167.47);
\definecolor{drawColor}{RGB}{0,0,0}

\path[draw=drawColor,line width= 0.4pt,line join=round,line cap=round] (253.93,123.01) --
	(254.59,121.35) --
	(255.26,119.76) --
	(255.93,118.25) --
	(256.60,116.80) --
	(257.27,115.42) --
	(257.93,114.10) --
	(258.60,112.84) --
	(259.27,111.64) --
	(259.94,110.49) --
	(260.61,109.39) --
	(261.27,108.34) --
	(261.94,107.34) --
	(262.61,106.39) --
	(263.28,105.48) --
	(263.95,104.61) --
	(264.61,103.78) --
	(265.28,102.99) --
	(265.95,102.23) --
	(266.62,101.51) --
	(267.28,100.82) --
	(267.95,100.17) --
	(268.62, 99.54) --
	(269.29, 98.95) --
	(269.96, 98.38) --
	(270.62, 97.84) --
	(271.29, 97.33) --
	(271.96, 96.84) --
	(272.63, 96.37) --
	(273.30, 95.93) --
	(273.96, 95.51) --
	(274.63, 95.11) --
	(275.30, 94.73) --
	(275.97, 94.36) --
	(276.63, 94.02) --
	(277.30, 93.70) --
	(277.97, 93.39) --
	(278.64, 93.10) --
	(279.31, 92.82) --
	(279.97, 92.56) --
	(280.64, 92.31) --
	(281.31, 92.08) --
	(281.98, 91.86) --
	(282.65, 91.66) --
	(283.31, 91.46) --
	(283.98, 91.28) --
	(284.65, 91.11) --
	(285.32, 90.95) --
	(285.99, 90.81) --
	(286.65, 90.67) --
	(287.32, 90.54) --
	(287.99, 90.43) --
	(288.66, 90.32) --
	(289.32, 90.22) --
	(289.99, 90.14) --
	(290.66, 90.06) --
	(291.33, 89.99) --
	(292.00, 89.92) --
	(292.66, 89.87) --
	(293.33, 89.82) --
	(294.00, 89.78) --
	(294.67, 89.75) --
	(295.34, 89.73) --
	(296.00, 89.71) --
	(296.67, 89.70) --
	(297.34, 89.70) --
	(298.01, 89.70) --
	(298.68, 89.71) --
	(299.34, 89.73) --
	(300.01, 89.76) --
	(300.68, 89.79) --
	(301.35, 89.82) --
	(302.01, 89.87) --
	(302.68, 89.92) --
	(303.35, 89.97) --
	(304.02, 90.03) --
	(304.69, 90.10) --
	(305.35, 90.17) --
	(306.02, 90.25) --
	(306.69, 90.34) --
	(307.36, 90.43) --
	(308.03, 90.53) --
	(308.69, 90.63) --
	(309.36, 90.74) --
	(310.03, 90.86) --
	(310.70, 90.98) --
	(311.37, 91.10) --
	(312.03, 91.24) --
	(312.70, 91.38) --
	(313.37, 91.52) --
	(314.04, 91.67) --
	(314.70, 91.83) --
	(315.37, 92.00) --
	(316.04, 92.17) --
	(316.71, 92.34) --
	(317.38, 92.53) --
	(318.04, 92.72) --
	(318.71, 92.92) --
	(319.38, 93.12) --
	(320.05, 93.33) --
	(320.72, 93.55) --
	(321.38, 93.77) --
	(322.05, 94.01) --
	(322.72, 94.25) --
	(323.39, 94.50) --
	(324.06, 94.75) --
	(324.72, 95.02) --
	(325.39, 95.29) --
	(326.06, 95.57) --
	(326.73, 95.86) --
	(327.39, 96.16) --
	(328.06, 96.46) --
	(328.73, 96.78) --
	(329.40, 97.11) --
	(330.07, 97.44) --
	(330.73, 97.79) --
	(331.40, 98.15) --
	(332.07, 98.51) --
	(332.74, 98.89) --
	(333.41, 99.28) --
	(334.07, 99.68) --
	(334.74,100.10) --
	(335.41,100.52) --
	(336.08,100.96) --
	(336.74,101.42) --
	(337.41,101.88) --
	(338.08,102.36) --
	(338.75,102.86) --
	(339.42,103.37) --
	(340.08,103.90) --
	(340.75,104.44) --
	(341.42,105.00) --
	(342.09,105.58) --
	(342.76,106.18) --
	(343.42,106.80) --
	(344.09,107.43) --
	(344.76,108.09) --
	(345.43,108.77) --
	(346.10,109.47) --
	(346.76,110.19) --
	(347.43,110.94) --
	(348.10,111.71) --
	(348.77,112.51) --
	(349.43,113.33) --
	(350.10,114.19) --
	(350.77,115.07) --
	(351.44,115.99) --
	(352.11,116.93) --
	(352.77,117.91) --
	(353.44,118.92) --
	(354.11,119.97) --
	(354.78,121.06) --
	(355.45,122.19) --
	(356.11,123.36) --
	(356.78,124.57) --
	(357.45,125.82) --
	(358.12,127.12) --
	(358.79,128.47) --
	(359.45,129.88) --
	(360.12,131.33) --
	(360.79,132.84) --
	(361.46,134.41) --
	(362.12,136.04) --
	(362.79,137.73) --
	(363.46,139.49) --
	(364.13,141.31) --
	(364.80,143.21) --
	(365.46,145.19) --
	(366.13,147.24) --
	(366.80,149.38) --
	(367.47,151.61) --
	(368.14,153.92) --
	(368.80,156.33) --
	(369.47,158.84) --
	(370.14,161.45) --
	(370.81,164.17) --
	(371.48,167.00) --
	(372.14,169.95) --
	(372.81,173.03) --
	(373.48,176.23) --
	(374.15,179.58) --
	(374.81,183.06) --
	(375.48,186.69) --
	(376.15,190.48) --
	(376.82,194.43) --
	(377.49,198.56) --
	(378.15,202.86) --
	(378.82,207.34) --
	(379.49,212.03) --
	(380.16,216.92) --
	(380.83,222.02) --
	(381.49,227.34) --
	(382.16,232.91) --
	(382.83,238.71) --
	(383.50,244.78) --
	(384.16,251.11) --
	(384.83,257.72) --
	(385.50,264.63);
\end{scope}
\begin{scope}
\path[clip] (  0.00,  0.00) rectangle (397.48,180.67);
\definecolor{drawColor}{RGB}{0,0,0}

\path[draw=drawColor,line width= 0.4pt,line join=round,line cap=round] (253.26, 54.00) -- (386.17, 54.00);

\path[draw=drawColor,line width= 0.4pt,line join=round,line cap=round] (253.26, 54.00) -- (253.26, 48.00);

\path[draw=drawColor,line width= 0.4pt,line join=round,line cap=round] (279.84, 54.00) -- (279.84, 48.00);

\path[draw=drawColor,line width= 0.4pt,line join=round,line cap=round] (306.42, 54.00) -- (306.42, 48.00);

\path[draw=drawColor,line width= 0.4pt,line join=round,line cap=round] (333.00, 54.00) -- (333.00, 48.00);

\path[draw=drawColor,line width= 0.4pt,line join=round,line cap=round] (359.59, 54.00) -- (359.59, 48.00);

\path[draw=drawColor,line width= 0.4pt,line join=round,line cap=round] (386.17, 54.00) -- (386.17, 48.00);

\node[text=drawColor,anchor=base,inner sep=0pt, outer sep=0pt, scale=  1.00] at (253.26, 32.40) {0.0};

\node[text=drawColor,anchor=base,inner sep=0pt, outer sep=0pt, scale=  1.00] at (279.84, 32.40) {0.2};

\node[text=drawColor,anchor=base,inner sep=0pt, outer sep=0pt, scale=  1.00] at (306.42, 32.40) {0.4};

\node[text=drawColor,anchor=base,inner sep=0pt, outer sep=0pt, scale=  1.00] at (333.00, 32.40) {0.6};

\node[text=drawColor,anchor=base,inner sep=0pt, outer sep=0pt, scale=  1.00] at (359.59, 32.40) {0.8};

\node[text=drawColor,anchor=base,inner sep=0pt, outer sep=0pt, scale=  1.00] at (386.17, 32.40) {1.0};

\path[draw=drawColor,line width= 0.4pt,line join=round,line cap=round] (247.94, 58.20) -- (247.94,163.27);

\path[draw=drawColor,line width= 0.4pt,line join=round,line cap=round] (247.94, 58.20) -- (241.94, 58.20);

\path[draw=drawColor,line width= 0.4pt,line join=round,line cap=round] (247.94, 84.47) -- (241.94, 84.47);

\path[draw=drawColor,line width= 0.4pt,line join=round,line cap=round] (247.94,110.74) -- (241.94,110.74);

\path[draw=drawColor,line width= 0.4pt,line join=round,line cap=round] (247.94,137.00) -- (241.94,137.00);

\path[draw=drawColor,line width= 0.4pt,line join=round,line cap=round] (247.94,163.27) -- (241.94,163.27);

\node[text=drawColor,rotate= 90.00,anchor=base,inner sep=0pt, outer sep=0pt, scale=  1.00] at (233.54, 58.20) {0};

\node[text=drawColor,rotate= 90.00,anchor=base,inner sep=0pt, outer sep=0pt, scale=  1.00] at (233.54, 84.47) {2};

\node[text=drawColor,rotate= 90.00,anchor=base,inner sep=0pt, outer sep=0pt, scale=  1.00] at (233.54,110.74) {4};

\node[text=drawColor,rotate= 90.00,anchor=base,inner sep=0pt, outer sep=0pt, scale=  1.00] at (233.54,137.00) {6};

\node[text=drawColor,rotate= 90.00,anchor=base,inner sep=0pt, outer sep=0pt, scale=  1.00] at (233.54,163.27) {8};
\end{scope}
\begin{scope}
\path[clip] (198.74,  0.00) rectangle (397.48,180.67);
\definecolor{drawColor}{RGB}{0,0,0}

\node[text=drawColor,anchor=base,inner sep=0pt, outer sep=0pt, scale=  1.00] at (319.71,  8.40) {Propensity};

\node[text=drawColor,rotate= 90.00,anchor=base,inner sep=0pt, outer sep=0pt, scale=  1.00] at (209.54,110.74) {Variance};
\end{scope}
\begin{scope}
\path[clip] (247.94, 54.00) rectangle (391.48,167.47);
\definecolor{drawColor}{RGB}{0,0,0}

\path[draw=drawColor,line width= 0.4pt,dash pattern=on 4pt off 4pt ,line join=round,line cap=round] (254.97,1770.61) --
	(255.26,1215.59) --
	(255.93,761.09) --
	(256.60,542.61) --
	(257.27,419.57) --
	(257.93,342.77) --
	(258.60,291.24) --
	(259.27,254.77) --
	(259.94,227.88) --
	(260.61,207.39) --
	(261.27,191.36) --
	(261.94,178.54) --
	(262.61,168.10) --
	(263.28,159.47) --
	(263.95,152.24) --
	(264.61,146.10) --
	(265.28,140.85) --
	(265.95,136.31) --
	(266.62,132.36) --
	(267.28,128.89) --
	(267.95,125.83) --
	(268.62,123.12) --
	(269.29,120.70) --
	(269.96,118.53) --
	(270.62,116.58) --
	(271.29,114.82) --
	(271.96,113.23) --
	(272.63,111.78) --
	(273.30,110.46) --
	(273.96,109.26) --
	(274.63,108.16) --
	(275.30,107.15) --
	(275.97,106.23) --
	(276.63,105.38) --
	(277.30,104.60) --
	(277.97,103.88) --
	(278.64,103.21) --
	(279.31,102.60) --
	(279.97,102.03) --
	(280.64,101.51) --
	(281.31,101.03) --
	(281.98,100.59) --
	(282.65,100.18) --
	(283.31, 99.80) --
	(283.98, 99.45) --
	(284.65, 99.13) --
	(285.32, 98.84) --
	(285.99, 98.57) --
	(286.65, 98.32) --
	(287.32, 98.09) --
	(287.99, 97.89) --
	(288.66, 97.71) --
	(289.32, 97.54) --
	(289.99, 97.39) --
	(290.66, 97.26) --
	(291.33, 97.14) --
	(292.00, 97.04) --
	(292.66, 96.96) --
	(293.33, 96.89) --
	(294.00, 96.83) --
	(294.67, 96.78) --
	(295.34, 96.75) --
	(296.00, 96.73) --
	(296.67, 96.72) --
	(297.34, 96.73) --
	(298.01, 96.74) --
	(298.68, 96.77) --
	(299.34, 96.80) --
	(300.01, 96.85) --
	(300.68, 96.91) --
	(301.35, 96.97) --
	(302.01, 97.05) --
	(302.68, 97.14) --
	(303.35, 97.23) --
	(304.02, 97.34) --
	(304.69, 97.45) --
	(305.35, 97.58) --
	(306.02, 97.71) --
	(306.69, 97.86) --
	(307.36, 98.01) --
	(308.03, 98.17) --
	(308.69, 98.34) --
	(309.36, 98.52) --
	(310.03, 98.71) --
	(310.70, 98.91) --
	(311.37, 99.11) --
	(312.03, 99.33) --
	(312.70, 99.56) --
	(313.37, 99.79) --
	(314.04,100.04) --
	(314.70,100.29) --
	(315.37,100.56) --
	(316.04,100.83) --
	(316.71,101.12) --
	(317.38,101.41) --
	(318.04,101.72) --
	(318.71,102.03) --
	(319.38,102.36) --
	(320.05,102.70) --
	(320.72,103.05) --
	(321.38,103.41) --
	(322.05,103.78) --
	(322.72,104.17) --
	(323.39,104.57) --
	(324.06,104.98) --
	(324.72,105.40) --
	(325.39,105.84) --
	(326.06,106.29) --
	(326.73,106.75) --
	(327.39,107.23) --
	(328.06,107.73) --
	(328.73,108.24) --
	(329.40,108.76) --
	(330.07,109.30) --
	(330.73,109.86) --
	(331.40,110.44) --
	(332.07,111.04) --
	(332.74,111.65) --
	(333.41,112.29) --
	(334.07,112.95) --
	(334.74,113.62) --
	(335.41,114.32) --
	(336.08,115.04) --
	(336.74,115.79) --
	(337.41,116.56) --
	(338.08,117.36) --
	(338.75,118.19) --
	(339.42,119.04) --
	(340.08,119.92) --
	(340.75,120.84) --
	(341.42,121.78) --
	(342.09,122.77) --
	(342.76,123.78) --
	(343.42,124.84) --
	(344.09,125.93) --
	(344.76,127.07) --
	(345.43,128.25) --
	(346.10,129.48) --
	(346.76,130.75) --
	(347.43,132.08) --
	(348.10,133.46) --
	(348.77,134.90) --
	(349.43,136.40) --
	(350.10,137.96) --
	(350.77,139.60) --
	(351.44,141.31) --
	(352.11,143.09) --
	(352.77,144.96) --
	(353.44,146.92) --
	(354.11,148.97) --
	(354.78,151.13) --
	(355.45,153.39) --
	(356.11,155.77) --
	(356.78,158.28) --
	(357.45,160.93) --
	(358.12,163.72) --
	(358.79,166.67) --
	(359.45,169.79) --
	(360.12,173.11) --
	(360.79,176.63) --
	(361.46,180.37) --
	(362.12,184.36) --
	(362.79,188.62) --
	(363.46,193.18) --
	(364.13,198.06) --
	(364.80,203.31) --
	(365.46,208.96) --
	(366.13,215.07) --
	(366.80,221.68) --
	(367.47,228.87) --
	(368.14,236.69) --
	(368.80,245.25) --
	(369.47,254.65) --
	(370.14,265.00) --
	(370.81,276.46) --
	(371.48,289.21) --
	(372.14,303.47) --
	(372.81,319.51) --
	(373.48,337.66) --
	(374.15,358.37) --
	(374.81,382.17) --
	(375.48,409.79) --
	(376.15,442.15) --
	(376.82,480.52) --
	(377.49,526.62) --
	(378.15,582.90) --
	(378.82,652.86) --
	(379.49,741.73) --
	(380.16,857.66) --
	(380.83,1013.87) --
	(381.49,1233.14) --
	(382.16,1557.76) --
	(382.44,1770.61);

\path[draw=drawColor,line width= 0.4pt,line join=round,line cap=round] (306.42, 81.19) rectangle (366.91, 59.59);

\path[draw=drawColor,line width= 0.4pt,line join=round,line cap=round] (311.82, 73.99) -- (322.62, 73.99);

\path[draw=drawColor,line width= 0.4pt,dash pattern=on 4pt off 4pt ,line join=round,line cap=round] (311.82, 66.79) -- (322.62, 66.79);

\node[text=drawColor,anchor=base west,inner sep=0pt, outer sep=0pt, scale=  0.60] at (328.02, 71.92) {Estimated-PS};

\node[text=drawColor,anchor=base west,inner sep=0pt, outer sep=0pt, scale=  0.60] at (328.02, 64.72) {True-PS};
\end{scope}
\end{tikzpicture}

\end{figure}
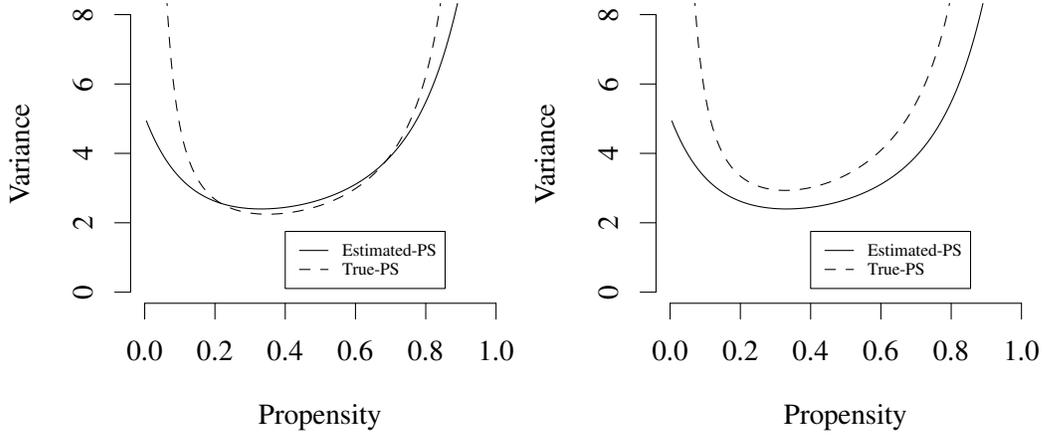

Figure \ref{comparison} illustrates these variances as a function of $p$ when $\sigma^2_1 =  4$ and $\sigma^2_0 = 16$. In the left panel we observe that when the data generating process has zero mean for both potential outcomes $\mu_1 = \mu_0 = 0$, the true propensity estimator can have lower variance for a certain range of $p$; the asymmetry in the plot is due to differing potential outcome variances, $\sigma_1^2 \neq \sigma_0^2$. The right panel shows that if either potential outcome surface is sufficiently far from zero, the quadratic term dominates and the estimated propensity IPW has lower variance than the true propensity IPW for all values of $p$. 

\section{Known propensity IPW as variable selection} \label{ipwvarsel}
Observe that for two strata, which we will denote $x_a$ and $x_b$, the true-propensity IPW effectively collapses cells whenever $p(x_a) = p(x_b)$:
\begin{equation}\label{sneaky_regression}
\begin{split}
\frac{N_a}{N} \bar{\tau}^{p, a} + \frac{N_b}{N} \bar{\tau}^{p, b} &= \frac{N_a}{N}\left( \frac{\hat{p}_a}{p} \bar{Y}_{a,1} - \frac{1-\hat{p}_a}{1-p} \bar{Y}_{a,0} \right) +  \frac{N_b}{N}\left( \frac{\hat{p}_b}{p} \bar{Y}_{b,1} - \frac{1-\hat{p}_b}{1-p} \bar{Y}_{b,0} \right) \\
& = \frac{N_a}{N}\left( \frac{N_{a,1}/N_a}{p} \frac{S_{a,1}}{N_{a,1}} - \frac{N_{a,0}/N_a}{p} \frac{S_{a,0}}{N_{a,0}} \right) + \\
& \qquad \frac{N_b}{N}\left( \frac{N_{a,1}/N_a}{p} \frac{S_{a,1}}{N_{a,1}} - \frac{N_{a,0}/N_a}{p} \frac{S_{a,0}}{N_{a,0}} \right)\\
&= \frac{1}{N} \left( \frac{S_{a,1} + S_{b,1}}{p} - \frac{S_{a,0} + S_{b,0}}{1 - p} \right )\\
& = \left( \frac{\hat{p}}{p} \bar{Y}_1 - \frac{1 - \hat{p}}{1 - p} \bar{Y}_0 \right )
\end{split}
\end{equation}
where, for notational ease, we write $N = N_a + N_b$ and $p = p(x_a) = p(x_b)$.  This demonstrates that, for discrete control variables, IPW estimators are fundamentally stratification estimators. 

Thus, consider an estimated-propensity IPW, but restricted to the unique values of $p(\mathcal{X})$ rather than all the unique levels of $\mathcal{X}$. That is, knowledge of the true propensity function may be used to define the strata without insisting that the finite-sample deviations in treatment assignment be ignored. Is such a ``hybrid'' IPW ever preferred over the estimated PS approach? Direct variance calculation reveals that, yes, the true propensity function can improve estimation in some cases. 

Here we demonstrate these calculations for two strata. The hybrid, collapsed strata, estimator has variance $\E(\sigma^2_1/N_1 + \sigma^2_0/N_0)$, while the conventional, uncollapsed, estimator has variance 
$$\frac{N_a^2}{N^2} \E\left(  \frac{\sigma^2_{a,1}}{N_{a,1}} + \frac{\sigma^2_{a,0}}{N_{a,0}}  \right) + \frac{N_b^2}{N^2} \E\left(  \frac{\sigma^2_{b,1}}{N_{b,1}} + \frac{\sigma^2_{b,0}}{N_{b,0}}  \right).$$ 
In general, which of these is smaller depends on the data generating process, but the collapsed estimator can certainly obtain lower variance if it omits extraneous controls. That is, consider the case where the means and variances for both potential outcomes are constant across the strata so that $\sigma^2_{a+b, z} = \sigma^2_{a, z} = \sigma^2_{b, z}$ for $z \in \{0, 1\}$ and further suppose that $N_a = N_b$ for simplicity. In that case, it can be shown that the hybrid (collapsing, or variable selecting) estimator has lower variance by showing that 
\begin{equation*}
\frac{\sigma^2_z}{4} \E\left(\frac{1}{N_{a,z} }+ \frac{1}{N_{b,z}}\right) = \frac{\sigma^2_z}{2} \E\left(\frac{1}{N_{a,z} }\right) \geq \sigma^2_z \E\left(\frac{1}{N_{a,z} + N_{b,z}}\right).
\end{equation*}

We first re-express this inequality so that the formulas from \cite{chao1972negative} may again be applied, this time for both $c = 1$ and $c = 2$. Let $N_a = N_b$. Because we require that each stratum has at least one treatment-control contrast, we have that $N_{a,1} = 1 + A$, where $A$ is a binomial random variable with parameter $n = N_a - 2$. Similarly $N_{b,1} = 1 + B$ for $B$ an independent random variable with the same distribution as $A$, because $p(x_a) = p(x_b)$ in the present example. Additionally, this implies that $N_{a + b, 1} = 2 + A + B$ for the same random variables, and we note that $A + B$ is a binomial random variable with parameter $N_a + N_b - 4 = 2n$. Therefore, showing that the collapsed estimator has lower variance amounts to showing that $$\E \left ( \frac{1}{2(1 + A)} \right )  - \E \left ( \frac{1}{2 + A + B}\right )  = \frac{1 - q^{(n+1)}}{2(n + 1)p} - \left ( \frac{1}{(2n+1)p}  - \frac{\left ( 1 - q^{(2n+2)} \right )}{(2n + 1)(2n+2)p^2} \right )$$ is nonnegative for all $p \in (0, 1)$, where $q = 1 - p$. We verify the nonnegativity of this difference in Appendix 1; Figure \ref{comparison2} illustrates the inequality. 

Note that, although the difference between the estimator variances appears small, it can be made arbitrarily large by considering additional extraneous stratification; these equations only compare the variance increase due to a single extraneous stratum.
\begin{figure}\caption{A comparison of the finite sample variances when $N_x = 17$, with $\sigma_1^2 = 4$ and $\sigma^2_0 = 16$ for the collapsed and ``non-collapsed" estimators.}\label{comparison2}
\centering

\begin{tikzpicture}[x=1pt,y=1pt]
\definecolor{fillColor}{RGB}{255,255,255}
\path[use as bounding box,fill opacity=0.00] (0,0) rectangle (216.81,180.67);
\begin{scope}
\path[clip] ( 49.20, 49.20) rectangle (203.61,167.47);
\definecolor{drawColor}{RGB}{0,0,0}

\path[draw=drawColor,line width= 0.4pt,dash pattern=on 1pt off 3pt ,line join=round,line cap=round] ( 55.64, 96.13) --
	( 56.36, 95.68) --
	( 57.07, 95.24) --
	( 57.79, 94.82) --
	( 58.51, 94.40) --
	( 59.23, 94.01) --
	( 59.95, 93.62) --
	( 60.67, 93.25) --
	( 61.38, 92.90) --
	( 62.10, 92.55) --
	( 62.82, 92.22) --
	( 63.54, 91.90) --
	( 64.26, 91.59) --
	( 64.98, 91.29) --
	( 65.70, 91.00) --
	( 66.41, 90.72) --
	( 67.13, 90.45) --
	( 67.85, 90.20) --
	( 68.57, 89.95) --
	( 69.29, 89.71) --
	( 70.01, 89.48) --
	( 70.72, 89.26) --
	( 71.44, 89.05) --
	( 72.16, 88.85) --
	( 72.88, 88.65) --
	( 73.60, 88.46) --
	( 74.32, 88.29) --
	( 75.04, 88.12) --
	( 75.75, 87.95) --
	( 76.47, 87.80) --
	( 77.19, 87.65) --
	( 77.91, 87.51) --
	( 78.63, 87.37) --
	( 79.35, 87.24) --
	( 80.06, 87.12) --
	( 80.78, 87.01) --
	( 81.50, 86.90) --
	( 82.22, 86.80) --
	( 82.94, 86.70) --
	( 83.66, 86.61) --
	( 84.38, 86.53) --
	( 85.09, 86.45) --
	( 85.81, 86.38) --
	( 86.53, 86.31) --
	( 87.25, 86.25) --
	( 87.97, 86.19) --
	( 88.69, 86.14) --
	( 89.40, 86.09) --
	( 90.12, 86.05) --
	( 90.84, 86.02) --
	( 91.56, 85.99) --
	( 92.28, 85.96) --
	( 93.00, 85.94) --
	( 93.72, 85.92) --
	( 94.43, 85.91) --
	( 95.15, 85.90) --
	( 95.87, 85.90) --
	( 96.59, 85.90) --
	( 97.31, 85.91) --
	( 98.03, 85.92) --
	( 98.74, 85.93) --
	( 99.46, 85.95) --
	(100.18, 85.97) --
	(100.90, 86.00) --
	(101.62, 86.03) --
	(102.34, 86.07) --
	(103.06, 86.11) --
	(103.77, 86.15) --
	(104.49, 86.20) --
	(105.21, 86.26) --
	(105.93, 86.31) --
	(106.65, 86.37) --
	(107.37, 86.44) --
	(108.08, 86.51) --
	(108.80, 86.58) --
	(109.52, 86.66) --
	(110.24, 86.74) --
	(110.96, 86.82) --
	(111.68, 86.91) --
	(112.40, 87.01) --
	(113.11, 87.10) --
	(113.83, 87.21) --
	(114.55, 87.31) --
	(115.27, 87.42) --
	(115.99, 87.54) --
	(116.71, 87.66) --
	(117.42, 87.78) --
	(118.14, 87.91) --
	(118.86, 88.04) --
	(119.58, 88.17) --
	(120.30, 88.31) --
	(121.02, 88.46) --
	(121.74, 88.60) --
	(122.45, 88.76) --
	(123.17, 88.92) --
	(123.89, 89.08) --
	(124.61, 89.25) --
	(125.33, 89.42) --
	(126.05, 89.59) --
	(126.76, 89.78) --
	(127.48, 89.96) --
	(128.20, 90.15) --
	(128.92, 90.35) --
	(129.64, 90.55) --
	(130.36, 90.76) --
	(131.07, 90.97) --
	(131.79, 91.19) --
	(132.51, 91.41) --
	(133.23, 91.64) --
	(133.95, 91.87) --
	(134.67, 92.11) --
	(135.39, 92.36) --
	(136.10, 92.61) --
	(136.82, 92.87) --
	(137.54, 93.13) --
	(138.26, 93.40) --
	(138.98, 93.68) --
	(139.70, 93.96) --
	(140.41, 94.25) --
	(141.13, 94.55) --
	(141.85, 94.85) --
	(142.57, 95.16) --
	(143.29, 95.48) --
	(144.01, 95.81) --
	(144.73, 96.14) --
	(145.44, 96.48) --
	(146.16, 96.83) --
	(146.88, 97.19) --
	(147.60, 97.56) --
	(148.32, 97.93) --
	(149.04, 98.31) --
	(149.75, 98.71) --
	(150.47, 99.11) --
	(151.19, 99.52) --
	(151.91, 99.94) --
	(152.63,100.37) --
	(153.35,100.81) --
	(154.07,101.26) --
	(154.78,101.72) --
	(155.50,102.20) --
	(156.22,102.68) --
	(156.94,103.18) --
	(157.66,103.68) --
	(158.38,104.20) --
	(159.09,104.73) --
	(159.81,105.28) --
	(160.53,105.84) --
	(161.25,106.41) --
	(161.97,106.99) --
	(162.69,107.59) --
	(163.41,108.21) --
	(164.12,108.84) --
	(164.84,109.48) --
	(165.56,110.14) --
	(166.28,110.82) --
	(167.00,111.51) --
	(167.72,112.23) --
	(168.43,112.95) --
	(169.15,113.70) --
	(169.87,114.47) --
	(170.59,115.25) --
	(171.31,116.06) --
	(172.03,116.89) --
	(172.75,117.73) --
	(173.46,118.60) --
	(174.18,119.49) --
	(174.90,120.41) --
	(175.62,121.34) --
	(176.34,122.31) --
	(177.06,123.29) --
	(177.77,124.30) --
	(178.49,125.34) --
	(179.21,126.41) --
	(179.93,127.51) --
	(180.65,128.63) --
	(181.37,129.78) --
	(182.09,130.97) --
	(182.80,132.18) --
	(183.52,133.43) --
	(184.24,134.71) --
	(184.96,136.03) --
	(185.68,137.38) --
	(186.40,138.77) --
	(187.11,140.20) --
	(187.83,141.66) --
	(188.55,143.17) --
	(189.27,144.72) --
	(189.99,146.31) --
	(190.71,147.94) --
	(191.43,149.62) --
	(192.14,151.35) --
	(192.86,153.12) --
	(193.58,154.94) --
	(194.30,156.82) --
	(195.02,158.74) --
	(195.74,160.72) --
	(196.45,162.76) --
	(197.17,164.85);
\end{scope}
\begin{scope}
\path[clip] (  0.00,  0.00) rectangle (216.81,180.67);
\definecolor{drawColor}{RGB}{0,0,0}

\path[draw=drawColor,line width= 0.4pt,line join=round,line cap=round] ( 54.92, 49.20) -- (197.89, 49.20);

\path[draw=drawColor,line width= 0.4pt,line join=round,line cap=round] ( 54.92, 49.20) -- ( 54.92, 43.20);

\path[draw=drawColor,line width= 0.4pt,line join=round,line cap=round] ( 83.51, 49.20) -- ( 83.51, 43.20);

\path[draw=drawColor,line width= 0.4pt,line join=round,line cap=round] (112.11, 49.20) -- (112.11, 43.20);

\path[draw=drawColor,line width= 0.4pt,line join=round,line cap=round] (140.70, 49.20) -- (140.70, 43.20);

\path[draw=drawColor,line width= 0.4pt,line join=round,line cap=round] (169.30, 49.20) -- (169.30, 43.20);

\path[draw=drawColor,line width= 0.4pt,line join=round,line cap=round] (197.89, 49.20) -- (197.89, 43.20);

\node[text=drawColor,anchor=base,inner sep=0pt, outer sep=0pt, scale=  1.00] at ( 54.92, 27.60) {0.0};

\node[text=drawColor,anchor=base,inner sep=0pt, outer sep=0pt, scale=  1.00] at ( 83.51, 27.60) {0.2};

\node[text=drawColor,anchor=base,inner sep=0pt, outer sep=0pt, scale=  1.00] at (112.11, 27.60) {0.4};

\node[text=drawColor,anchor=base,inner sep=0pt, outer sep=0pt, scale=  1.00] at (140.70, 27.60) {0.6};

\node[text=drawColor,anchor=base,inner sep=0pt, outer sep=0pt, scale=  1.00] at (169.30, 27.60) {0.8};

\node[text=drawColor,anchor=base,inner sep=0pt, outer sep=0pt, scale=  1.00] at (197.89, 27.60) {1.0};

\path[draw=drawColor,line width= 0.4pt,line join=round,line cap=round] ( 49.20, 53.58) -- ( 49.20,163.09);

\path[draw=drawColor,line width= 0.4pt,line join=round,line cap=round] ( 49.20, 53.58) -- ( 43.20, 53.58);

\path[draw=drawColor,line width= 0.4pt,line join=round,line cap=round] ( 49.20, 80.96) -- ( 43.20, 80.96);

\path[draw=drawColor,line width= 0.4pt,line join=round,line cap=round] ( 49.20,108.34) -- ( 43.20,108.34);

\path[draw=drawColor,line width= 0.4pt,line join=round,line cap=round] ( 49.20,135.72) -- ( 43.20,135.72);

\path[draw=drawColor,line width= 0.4pt,line join=round,line cap=round] ( 49.20,163.09) -- ( 43.20,163.09);

\node[text=drawColor,rotate= 90.00,anchor=base,inner sep=0pt, outer sep=0pt, scale=  1.00] at ( 34.80, 53.58) {0};

\node[text=drawColor,rotate= 90.00,anchor=base,inner sep=0pt, outer sep=0pt, scale=  1.00] at ( 34.80, 80.96) {2};

\node[text=drawColor,rotate= 90.00,anchor=base,inner sep=0pt, outer sep=0pt, scale=  1.00] at ( 34.80,108.34) {4};

\node[text=drawColor,rotate= 90.00,anchor=base,inner sep=0pt, outer sep=0pt, scale=  1.00] at ( 34.80,135.72) {6};

\node[text=drawColor,rotate= 90.00,anchor=base,inner sep=0pt, outer sep=0pt, scale=  1.00] at ( 34.80,163.09) {8};
\end{scope}
\begin{scope}
\path[clip] (  0.00,  0.00) rectangle (216.81,180.67);
\definecolor{drawColor}{RGB}{0,0,0}

\node[text=drawColor,anchor=base,inner sep=0pt, outer sep=0pt, scale=  1.00] at (126.41,  3.60) {Propensity};

\node[text=drawColor,rotate= 90.00,anchor=base,inner sep=0pt, outer sep=0pt, scale=  1.00] at ( 10.80,108.34) {Variance};
\end{scope}
\begin{scope}
\path[clip] ( 49.20, 49.20) rectangle (203.61,167.47);
\definecolor{drawColor}{RGB}{0,0,0}

\path[draw=drawColor,line width= 0.4pt,line join=round,line cap=round] ( 55.64, 96.27) --
	( 56.36, 95.95) --
	( 57.07, 95.64) --
	( 57.79, 95.33) --
	( 58.51, 95.03) --
	( 59.23, 94.73) --
	( 59.95, 94.45) --
	( 60.67, 94.17) --
	( 61.38, 93.90) --
	( 62.10, 93.63) --
	( 62.82, 93.38) --
	( 63.54, 93.12) --
	( 64.26, 92.88) --
	( 64.98, 92.64) --
	( 65.70, 92.41) --
	( 66.41, 92.19) --
	( 67.13, 91.97) --
	( 67.85, 91.75) --
	( 68.57, 91.55) --
	( 69.29, 91.35) --
	( 70.01, 91.16) --
	( 70.72, 90.97) --
	( 71.44, 90.79) --
	( 72.16, 90.61) --
	( 72.88, 90.44) --
	( 73.60, 90.28) --
	( 74.32, 90.12) --
	( 75.04, 89.97) --
	( 75.75, 89.82) --
	( 76.47, 89.68) --
	( 77.19, 89.55) --
	( 77.91, 89.42) --
	( 78.63, 89.29) --
	( 79.35, 89.17) --
	( 80.06, 89.06) --
	( 80.78, 88.95) --
	( 81.50, 88.85) --
	( 82.22, 88.75) --
	( 82.94, 88.66) --
	( 83.66, 88.57) --
	( 84.38, 88.49) --
	( 85.09, 88.42) --
	( 85.81, 88.34) --
	( 86.53, 88.28) --
	( 87.25, 88.22) --
	( 87.97, 88.16) --
	( 88.69, 88.11) --
	( 89.40, 88.06) --
	( 90.12, 88.02) --
	( 90.84, 87.99) --
	( 91.56, 87.96) --
	( 92.28, 87.93) --
	( 93.00, 87.91) --
	( 93.72, 87.89) --
	( 94.43, 87.88) --
	( 95.15, 87.87) --
	( 95.87, 87.87) --
	( 96.59, 87.87) --
	( 97.31, 87.88) --
	( 98.03, 87.89) --
	( 98.74, 87.91) --
	( 99.46, 87.93) --
	(100.18, 87.96) --
	(100.90, 87.99) --
	(101.62, 88.03) --
	(102.34, 88.07) --
	(103.06, 88.12) --
	(103.77, 88.17) --
	(104.49, 88.22) --
	(105.21, 88.28) --
	(105.93, 88.35) --
	(106.65, 88.42) --
	(107.37, 88.49) --
	(108.08, 88.57) --
	(108.80, 88.66) --
	(109.52, 88.75) --
	(110.24, 88.85) --
	(110.96, 88.95) --
	(111.68, 89.05) --
	(112.40, 89.16) --
	(113.11, 89.28) --
	(113.83, 89.40) --
	(114.55, 89.52) --
	(115.27, 89.65) --
	(115.99, 89.79) --
	(116.71, 89.93) --
	(117.42, 90.08) --
	(118.14, 90.23) --
	(118.86, 90.39) --
	(119.58, 90.55) --
	(120.30, 90.72) --
	(121.02, 90.90) --
	(121.74, 91.08) --
	(122.45, 91.26) --
	(123.17, 91.45) --
	(123.89, 91.65) --
	(124.61, 91.85) --
	(125.33, 92.06) --
	(126.05, 92.28) --
	(126.76, 92.50) --
	(127.48, 92.73) --
	(128.20, 92.96) --
	(128.92, 93.20) --
	(129.64, 93.45) --
	(130.36, 93.70) --
	(131.07, 93.96) --
	(131.79, 94.22) --
	(132.51, 94.50) --
	(133.23, 94.77) --
	(133.95, 95.06) --
	(134.67, 95.35) --
	(135.39, 95.65) --
	(136.10, 95.96) --
	(136.82, 96.28) --
	(137.54, 96.60) --
	(138.26, 96.93) --
	(138.98, 97.26) --
	(139.70, 97.61) --
	(140.41, 97.96) --
	(141.13, 98.32) --
	(141.85, 98.69) --
	(142.57, 99.07) --
	(143.29, 99.45) --
	(144.01, 99.85) --
	(144.73,100.25) --
	(145.44,100.66) --
	(146.16,101.08) --
	(146.88,101.51) --
	(147.60,101.95) --
	(148.32,102.39) --
	(149.04,102.85) --
	(149.75,103.32) --
	(150.47,103.79) --
	(151.19,104.28) --
	(151.91,104.77) --
	(152.63,105.28) --
	(153.35,105.79) --
	(154.07,106.32) --
	(154.78,106.86) --
	(155.50,107.41) --
	(156.22,107.97) --
	(156.94,108.54) --
	(157.66,109.12) --
	(158.38,109.71) --
	(159.09,110.32) --
	(159.81,110.93) --
	(160.53,111.56) --
	(161.25,112.20) --
	(161.97,112.86) --
	(162.69,113.52) --
	(163.41,114.20) --
	(164.12,114.89) --
	(164.84,115.60) --
	(165.56,116.32) --
	(166.28,117.05) --
	(167.00,117.80) --
	(167.72,118.56) --
	(168.43,119.34) --
	(169.15,120.13) --
	(169.87,120.93) --
	(170.59,121.76) --
	(171.31,122.59) --
	(172.03,123.44) --
	(172.75,124.31) --
	(173.46,125.20) --
	(174.18,126.10) --
	(174.90,127.01) --
	(175.62,127.95) --
	(176.34,128.90) --
	(177.06,129.87) --
	(177.77,130.86) --
	(178.49,131.86) --
	(179.21,132.89) --
	(179.93,133.93) --
	(180.65,134.99) --
	(181.37,136.07) --
	(182.09,137.17) --
	(182.80,138.29) --
	(183.52,139.43) --
	(184.24,140.59) --
	(184.96,141.77) --
	(185.68,142.98) --
	(186.40,144.20) --
	(187.11,145.45) --
	(187.83,146.72) --
	(188.55,148.01) --
	(189.27,149.32) --
	(189.99,150.66) --
	(190.71,152.02) --
	(191.43,153.40) --
	(192.14,154.81) --
	(192.86,156.25) --
	(193.58,157.70) --
	(194.30,159.19) --
	(195.02,160.70) --
	(195.74,162.24) --
	(196.45,163.80) --
	(197.17,165.39);

\path[draw=drawColor,line width= 0.4pt,dash pattern=on 1pt off 3pt ,line join=round,line cap=round] ( 55.64, 96.13) --
	( 56.36, 95.68) --
	( 57.07, 95.24) --
	( 57.79, 94.82) --
	( 58.51, 94.40) --
	( 59.23, 94.01) --
	( 59.95, 93.62) --
	( 60.67, 93.25) --
	( 61.38, 92.90) --
	( 62.10, 92.55) --
	( 62.82, 92.22) --
	( 63.54, 91.90) --
	( 64.26, 91.59) --
	( 64.98, 91.29) --
	( 65.70, 91.00) --
	( 66.41, 90.72) --
	( 67.13, 90.45) --
	( 67.85, 90.20) --
	( 68.57, 89.95) --
	( 69.29, 89.71) --
	( 70.01, 89.48) --
	( 70.72, 89.26) --
	( 71.44, 89.05) --
	( 72.16, 88.85) --
	( 72.88, 88.65) --
	( 73.60, 88.46) --
	( 74.32, 88.29) --
	( 75.04, 88.12) --
	( 75.75, 87.95) --
	( 76.47, 87.80) --
	( 77.19, 87.65) --
	( 77.91, 87.51) --
	( 78.63, 87.37) --
	( 79.35, 87.24) --
	( 80.06, 87.12) --
	( 80.78, 87.01) --
	( 81.50, 86.90) --
	( 82.22, 86.80) --
	( 82.94, 86.70) --
	( 83.66, 86.61) --
	( 84.38, 86.53) --
	( 85.09, 86.45) --
	( 85.81, 86.38) --
	( 86.53, 86.31) --
	( 87.25, 86.25) --
	( 87.97, 86.19) --
	( 88.69, 86.14) --
	( 89.40, 86.09) --
	( 90.12, 86.05) --
	( 90.84, 86.02) --
	( 91.56, 85.99) --
	( 92.28, 85.96) --
	( 93.00, 85.94) --
	( 93.72, 85.92) --
	( 94.43, 85.91) --
	( 95.15, 85.90) --
	( 95.87, 85.90) --
	( 96.59, 85.90) --
	( 97.31, 85.91) --
	( 98.03, 85.92) --
	( 98.74, 85.93) --
	( 99.46, 85.95) --
	(100.18, 85.97) --
	(100.90, 86.00) --
	(101.62, 86.03) --
	(102.34, 86.07) --
	(103.06, 86.11) --
	(103.77, 86.15) --
	(104.49, 86.20) --
	(105.21, 86.26) --
	(105.93, 86.31) --
	(106.65, 86.37) --
	(107.37, 86.44) --
	(108.08, 86.51) --
	(108.80, 86.58) --
	(109.52, 86.66) --
	(110.24, 86.74) --
	(110.96, 86.82) --
	(111.68, 86.91) --
	(112.40, 87.01) --
	(113.11, 87.10) --
	(113.83, 87.21) --
	(114.55, 87.31) --
	(115.27, 87.42) --
	(115.99, 87.54) --
	(116.71, 87.66) --
	(117.42, 87.78) --
	(118.14, 87.91) --
	(118.86, 88.04) --
	(119.58, 88.17) --
	(120.30, 88.31) --
	(121.02, 88.46) --
	(121.74, 88.60) --
	(122.45, 88.76) --
	(123.17, 88.92) --
	(123.89, 89.08) --
	(124.61, 89.25) --
	(125.33, 89.42) --
	(126.05, 89.59) --
	(126.76, 89.78) --
	(127.48, 89.96) --
	(128.20, 90.15) --
	(128.92, 90.35) --
	(129.64, 90.55) --
	(130.36, 90.76) --
	(131.07, 90.97) --
	(131.79, 91.19) --
	(132.51, 91.41) --
	(133.23, 91.64) --
	(133.95, 91.87) --
	(134.67, 92.11) --
	(135.39, 92.36) --
	(136.10, 92.61) --
	(136.82, 92.87) --
	(137.54, 93.13) --
	(138.26, 93.40) --
	(138.98, 93.68) --
	(139.70, 93.96) --
	(140.41, 94.25) --
	(141.13, 94.55) --
	(141.85, 94.85) --
	(142.57, 95.16) --
	(143.29, 95.48) --
	(144.01, 95.81) --
	(144.73, 96.14) --
	(145.44, 96.48) --
	(146.16, 96.83) --
	(146.88, 97.19) --
	(147.60, 97.56) --
	(148.32, 97.93) --
	(149.04, 98.31) --
	(149.75, 98.71) --
	(150.47, 99.11) --
	(151.19, 99.52) --
	(151.91, 99.94) --
	(152.63,100.37) --
	(153.35,100.81) --
	(154.07,101.26) --
	(154.78,101.72) --
	(155.50,102.20) --
	(156.22,102.68) --
	(156.94,103.18) --
	(157.66,103.68) --
	(158.38,104.20) --
	(159.09,104.73) --
	(159.81,105.28) --
	(160.53,105.84) --
	(161.25,106.41) --
	(161.97,106.99) --
	(162.69,107.59) --
	(163.41,108.21) --
	(164.12,108.84) --
	(164.84,109.48) --
	(165.56,110.14) --
	(166.28,110.82) --
	(167.00,111.51) --
	(167.72,112.23) --
	(168.43,112.95) --
	(169.15,113.70) --
	(169.87,114.47) --
	(170.59,115.25) --
	(171.31,116.06) --
	(172.03,116.89) --
	(172.75,117.73) --
	(173.46,118.60) --
	(174.18,119.49) --
	(174.90,120.41) --
	(175.62,121.34) --
	(176.34,122.31) --
	(177.06,123.29) --
	(177.77,124.30) --
	(178.49,125.34) --
	(179.21,126.41) --
	(179.93,127.51) --
	(180.65,128.63) --
	(181.37,129.78) --
	(182.09,130.97) --
	(182.80,132.18) --
	(183.52,133.43) --
	(184.24,134.71) --
	(184.96,136.03) --
	(185.68,137.38) --
	(186.40,138.77) --
	(187.11,140.20) --
	(187.83,141.66) --
	(188.55,143.17) --
	(189.27,144.72) --
	(189.99,146.31) --
	(190.71,147.94) --
	(191.43,149.62) --
	(192.14,151.35) --
	(192.86,153.12) --
	(193.58,154.94) --
	(194.30,156.82) --
	(195.02,158.74) --
	(195.74,160.72) --
	(196.45,162.76) --
	(197.17,164.85);

\path[draw=drawColor,line width= 0.4pt,line join=round,line cap=round] (119.26, 80.96) rectangle (190.20, 57.56);

\path[draw=drawColor,line width= 0.4pt,line join=round,line cap=round] (125.11, 73.16) -- (136.81, 73.16);

\path[draw=drawColor,line width= 0.4pt,dash pattern=on 1pt off 3pt ,line join=round,line cap=round] (125.11, 65.36) -- (136.81, 65.36);

\node[text=drawColor,anchor=base west,inner sep=0pt, outer sep=0pt, scale=  0.65] at (142.66, 70.92) {Estimated IPW};

\node[text=drawColor,anchor=base west,inner sep=0pt, outer sep=0pt, scale=  0.65] at (142.66, 63.12) {Collapsed IPW};
\end{scope}
\end{tikzpicture}

\end{figure}
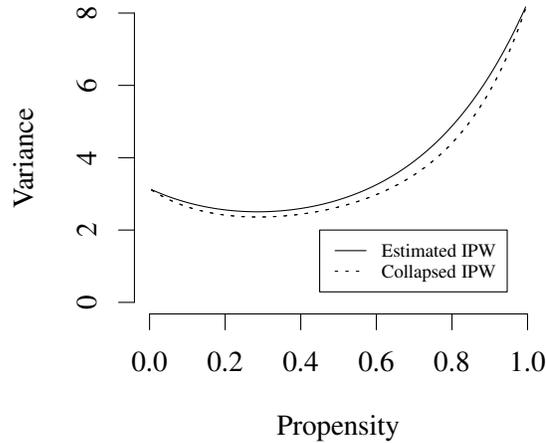
Finally, we note that a known propensity function does not discern {\em instruments} --- variables that only impact the outcome via their causal effect on treatment assignment. But the above variance comparisons equally as well apply to comparing between two distinct valid propensity functions, one with instruments and one without. By definition, the instrumental strata have identical potential outcome means and variances and so are ``noise'' variables in the relevant sense.

\newpage
\section{Discussion}
The irrelevance of a known propensity function is essentially a quirk of IPW estimators. Finite sample variance calculations highlight that the true propensity IPW is a perverse estimator which has variance that increases unboundedly as the data are shifted away from the origin. However, a known propensity function can be advantageous if it permits screening out unnecessary controls. This observation opens the door for auxiliary data to be used to estimate a propensity function for this purpose, a natural application of semi-supervised learning \citep{liang2007use} to treatment effect estimation.

Notably, an estimated-propensity IPW can still outperform a variable-selected IPW if the true propensity score ends up dropping valuable prognostic factors. That is, ``estimating the propensity scores'' using variables known to be irrelevant for treatment assignment can nonetheless increase statistical precision; in such cases the IPW estimator is acting as a direct regression adjustment in disguise.

\bibliographystyle{biometrika}
\bibliography{propensity-scores}

\appendix

\appendixone
\section*{Appendix 1}
Here we give a proof that omitting noise variables from an IPW adjustment reduces variance, as discussed in Section \ref{ipwvarsel}. Specifically, we will demonstrate that $$\frac{1 - q^{(n+1)}}{2(n + 1)p} - \left ( \frac{1}{(2n+1)p}  - \frac{\left ( 1 - q^{(2n+2)} \right )}{(2n + 1)(2n+2)p^2} \right ) \geq 0.$$ 

To begin, observe that
\begin{equation*}
\begin{aligned}
\frac{1 - q^{(n+1)}}{2(n + 1)p} - \left ( \frac{1}{(2n+1)p}  - \frac{\left ( 1 - q^{(2n+2)} \right )}{(2n + 1)(2n+2)p^2} \right ) &= 
\frac{1 - q^{(n+1)}}{2(n + 1)p} - \frac{1}{(2n+1)p} + \frac{\left ( 1 - q^{(2(n+1))} \right )}{(2n + 1)(2(n+1))p^2}\\
&= \frac{(1 - q^{(n+1)})(2n+1)p - 2(n + 1)p + (1 - q^{(2(n+1))})}{2(n + 1)(2n+1)p^2}
\end{aligned}
\end{equation*}
and that $2(n + 1)(2n+1)p^2 \geq 0$ for all $p \in [0, 1]$ and $n \geq 1$.
Thus we aim to show that 
$$(1 - q^{(n+1)})(2n+1)(1-q) - 2(n + 1)(1-q) + (1 - q^{(2(n+1))}) \geq 0,$$
which simplifies to $$q(1 - q^{(n+1)})(q^{n} + 1) - (2n+2)(1-q)q^{(n+1)} \geq 0,$$ the lefthand side of which is $0$ at both $q = 0$ and $q = 1$. Because $q \geq 0$, dividing by $q$ further reduces our task to showing that
$$(1 - q^{(n+1)})(q^{n} + 1) - (2n+2)(1-q)q^n \geq 0.$$
Here, the lefthand side is $0$ at $p = 0$ (i.e. $q = 1$) and $1$ at $p = 1$ (i.e. $q = 0$); therefore, we must show that this expression is nondecreasing in $p$, which we do by evaluating its derivative:
\begin{equation*}
\begin{aligned}
\frac{d}{dp} \left\{(1 - q^{(n+1)})(q^{n} + 1) - (2n+2)(1-q)q^n\right\} %&= \frac{d}{dq} \left\{(1 - q^{(n+1)})(q^{n} + 1) - (2n+2)(1-q)q^n\right\} \frac{dq}{dp}\\
&= -\frac{d}{dq} \left\{(1 - q^{(n+1)})(q^{n} + 1) - (2n+2)(1-q)q^n\right\}\\
&= - (2n+1)q^{n-1}\left[ n(q-1) - q(q^n - 1) \right].
\end{aligned}
\end{equation*}
Following a similar strategy to before, we factor out $(2n+1)q^{n-1}$, which is nonnegative in $p$, and aim to show that $- \left[ n(q-1) - q(q^n - 1) \right]$ is nonnegative for all $p \in [0, 1]$.  As before, observe that $- \left[ n(q-1) - q(q^n - 1) \right] = 0$ when $p = 0$ and $n$ when $p = 1$, meaning we must now show that {\em this} expression is nondecreasing in $p$. Again taking the derivative with respect to $p$, we find that
\begin{equation*}
\begin{aligned}
\frac{d}{dp} - \left[ n(q-1) - q(q^n - 1) \right] &= \frac{d}{dq} \left[ n(q-1) - q(q^n - 1) \right]\\
&= (n+1)(1-q^n) = (n+1)(1-(1-p)^n),
\end{aligned}
\end{equation*}
which is nonnegative for all $p \in [0,1]$.

\end{document}